\documentclass{article}

\begin{document}

\def\R{{\bf R}}

\title{Fuchsian methods and spacetime singularities}
\author{Alan D. Rendall}

\maketitle

\begin{abstract}
Fuchsian methods and their applications to the study of the structure of
spacetime singularities are surveyed. The existence question for spacetimes
with compact Cauchy horizons is discussed. After some basic facts concerning 
Fuchsian equations have been recalled, various ways in which these equations
have been applied in general relativity are described. Possible future 
applications are indicated.  
\end{abstract}

\section{Introduction}\label{intro}

The singularity theorems of Penrose and Hawking showed that very often
something goes wrong with the evolution of spacetimes according to the
Einstein equations. They did not say what goes wrong and more than
thirty years later we still have only limited information concerning
the structure of general spacetime singularities. This situation
represents an outstanding challenge to mathematical relativists. The most
general rigorous results we possess on this subject have been obtained 
by Fuchsian methods. The purpose of this article is to give a portrait
of the present state of development of the applications of these methods 
to general relativity. It also gives some indications as to what future
improvements may be anticipated in this area.

How can we learn about singular solutions of differential equations?
By looking at explicit solutions it is often possible to make a
guess as to what the singularities might be like in general. In the 
case of equations arising in physics or other parts of science intuition
coming from the applications may also make a contribution. Trial
and error should not be neglected either. Once a conjectural
picture of singularities of solutions of a particular equation
has developed the next step is to confront it with the equation and 
to see if it is formally consistent. When formal consistency holds
this is enough to make some consider the problem to be 
solved. Others see it as evidence for a certain solution of the
problem but still want to have a definitive confirmation which
can only be provided by rigorous mathematical theorems. Fuchsian
methods allow the step from an intuitive to a rigorous result to
be taken in many cases. When they apply they provide statements 
of the form that to each formal solution there corresponds a unique 
solution of the exact equation.  

In the case of cosmological singularities in solutions of the
Einstein equations the most influential heuristic work is that of
Belinskii, Khalatnikov and Lifshitz (BKL). One important element
of the picture they developed is the idea that near the singularity
the dynamics at different spatial points should decouple. Thus
the partial differential equations describing self-gravitating
matter decouple asymptotically in the limit and reduce to ordinary
differential equations. Furthermore, for many matter fields and
assuming four spacetime dimensions, the limiting dynamics of generic
solutions is that of a family of solutions of the Bianchi type IX 
vacuum equations (mixmaster model) with the spatial coordinates acting 
as parameters. The solutions of the mixmaster model have a complicated 
oscillatory behaviour and it is only recently that the principal 
features of their dynamics were understood rigorously \cite{ringstrom00},
\cite{ringstrom01}. In view of this fact it is not surprising that the 
role of the mixmaster model as a template for the evolution of spatially 
inhomogeneous solutions of the Einstein equations has not been captured 
in mathematical theorems. It has been clearly exhibited in numerical 
calculations \cite{berger01b}.

The BKL picture also predicts for which dimensions and combinations
of matter fields oscillatory behaviour is not to be expected. In
such cases the dynamics suggested by the BKL analysis is much
simpler, with the most important geometric quantities varying
in a monotone fashion near the singularity. It is in this situation
that Fuchsian techniques can be expected to apply and indeed this
has turned out to be the case in a variety of examples.

The paper is organized as follows. Section \ref{horizon} describes
the earliest application of Fuchsian techniques to general relativity
known to the author, namely the construction of large classes of
vacuum spacetimes with Cauchy horizons. In Section \ref{kr} a
general introduction to Fuchsian techniques is given and an
existence and uniqueness theorem of wide applicability is presented.
A number of applications of this theorem to the construction of classes 
of spacetimes with certain kinds of singularities are described in
Section \ref{examples}. The last section lists existing generalizations
of these results and suggests potential further developments. 

\section{Cauchy horizons}\label{horizon}

One of the most interesting issues concerning the nature of spacetime
singularites is that of the strong cosmic censorship hypothesis. We
recall the formulation of this hypothesis given by Eardley and Moncrief
\cite{eardley}. Corresponding to any initial data set for the 
Einstein-matter equations on a compact manifold there is a unique
maximal Cauchy development which, intuitively, is the largest globally
hyperbolic solution of the Einstein equations with the given initial 
data. The formulation of the strong cosmic censorship hypothesis which
is of interest here is that for generic initial data on the given 
manifold the maximal Cauchy development should be inextendible. This
means that for almost all initial data it is not possible to embed the
maximal Cauchy development in a bigger solution of the Einstein-matter
equations for which the hypersurface carrying the initial data remains 
a Cauchy surface. If this failed the original initial hypersurface would
not be a Cauchy surface for the extended spacetime in which the maximal
Cauchy development was embedded. This means that the uniqueness of the
time development in terms of initial data would no longer be guaranteed 
and that, from a physical point of view, predictability would break down.

In order to make the above formulation complete it is necessary to
fix the notion of generic initial data precisely. Usually this is
done by requiring this set of initial data to contain an open dense
subset of the set of all initial data on the given manifold in a
suitable topology. There are obvious candidates for the topology
to be used. More problematic is the question of the choice of
matter fields. As formulated the hypothesis depends implicitly on
fixing a particular type of matter fields to describe the matter 
content of spacetime. In fact we cannot expect it to be true without
limitations on the kind of matter chosen. One criterion is that the 
given matter model develops no singularities when considered as a
test field in Minkowski space. Examples are a linear scalar field,
the Maxwell field or collisionless matter described by the Vlasov 
equation. (See \cite{rendall02a} for a general discussion of the
Einstein-Vlasov system.) Examples of matter models which form 
singularities in flat space and which are therefore bad from the point of 
view of cosmic censorship include dust and perfect fluids with pressure.
One way of avoiding the issue of the choice of matter fields is to
concentrate on the vacuum case. 
  
When it does happen that the maximal Cauchy development of certain 
initial data is extendible the boundary of the maximal Cauchy development
in the extension is by definition a Cauchy horizon. It is null 
hypersurface with some (perhaps low) degree of regularity. A
possible approach to learning more about cosmic censorship is to
construct spacetimes which contain a Cauchy horizon and are otherwise
as general as possible. A clean case to start with is that where the
Cauchy horizon is compact and smooth.

In \cite{moncrief82} and \cite{moncrief84} Moncrief proved the existence of 
large classes
of vacuum spacetimes with smooth compact Cauchy horizons. All these spacetimes
have one spacelike Killing vector and are analytic ($C^\omega$). He did
so by solving a Cauchy problem for a class of singular partial differential
equations. This is apparently the first occurrence in the literature on
general relativity of Fuchsian methods, a technique which will be introduced 
in the next section. The presence of a Killing vector is not an 
accident and Moncrief and Isenberg \cite{moncrief83} have shown that 
under suitable conditions the existence of a compact analytic Cauchy
horizon automatically implies the existence of at least one Killing vector.
Later it was found that analyticity could be replaced by the weaker
condition of smoothness in the conditions of their theorem 
\cite{friedrich99}. Moreover it has been shown that in this context 
a compact Cauchy horizon in a smooth spacetime is itself smooth 
\cite{chrusciel01}. Moncrief and Isenberg conjectured
in \cite{isenberg92} that any vacuum spacetime with a smooth compact 
Cauchy horizon has a Killing vector which is tangent to the generators
of the horizon and whose integral curves are closed. This means that
the Cauchy horizon, considered as an abstract manifold, admits an
action of the circle $S^1$. Not all three-dimensional manifolds admit
such an action and thus a restriction on the topology of the manifold
results. It must be a Seifert manifold. All these manifolds admit
geometric structures in the sense of Thurston and six of the Thurston
geometries can be realized in this way. In \cite{rendall98} it was shown 
by a quite different method, using the Cheeger-Gromov theory of
collapsing Riemannian manifolds with bounded curvature, that a
smooth compact Cauchy horizon in a spacetime satifying the Einstein
equations coupled to certain matter models has a topology which is
restricted to those which admit one of seven Thurston
geometries. These are the six corresponding to Seifert manifolds
together with the type Sol. In terminology more familiar to relativists
this last type is related to Bianchi type VI${}_0$. In \cite{rendall98} it 
was shown that all topologies of Seifert type allowed by this classification 
occur in locally spatially homogeneous solutions of the vacuum Einstein 
equations.

The results of Moncrief and Isenberg are limited to the case of the
vacuum Einstein equations. It is interesting to ask to what extent
similar results are true in the presence of matter. Some insight into this 
question can be obtained by looking at the spatially homogeneous case.
There are (locally) homogeneous vacuum spacetimes which have a compact
Cauchy surface and a compact Cauchy horizon (cf. \cite{chrusciel95}.) In
contrast to this, locally spatially homogeneous spacetimes with a
compact Cauchy surface and phenomenological matter can be shown to
have no Cauchy horizon in many cases \cite{rendall95}. This suggests that
the generality of the spacetimes containing Cauchy horizons produced by
Moncrief's construction is essentially dependent on the absence of matter
sources in the field equations. If this is true then the Isenberg-Moncrief
procedure is also limited to the source-free case. (Here the term source-free
is meant to include the case of a source-free Maxwell field which acts as a 
source in the Einstein equations.) 

\section{A general existence and uniqueness theorem}\label{kr}

This section describes the theory of Fuchsian equations. This class of
equations has the form
\begin{equation}\label{fuchs}
t\partial u/\partial t+Nu=tf(t,x,u,u_x)
\end{equation}
Here $u$ is a function of a real number $t$, thought of as a time coordinate,
and a point $x$ in $\R^n$, thought of as a point of space. This function
takes values in $\R^k$. The $k\times k$ matrix-valued function $N$
depends only on $x$. It satisfies some positivity condition, as will
be discussed in more detail later. The function $f$ satisfies some 
regularity conditions which will also be specified later. The notation
$u_x$ is used as a shorthand for the collection of partial derivatives
of $u$ with respect to the variables $x$. For applications it is useful
to note the following. Suppose that instead of (\ref{fuchs}) we had the
equation
\begin{equation}\label{fuchsprime}
t\partial u/\partial t+Nu=t^\alpha f(t,x,u,u_x)
\end{equation}
for some positive real number $\alpha$. Defining $s=t^\alpha$ leads to
the equation
\begin{equation}\label{fuchsdoubleprime}
s\partial u/\partial s+\alpha Nu=s [\alpha f(s^{1/\alpha},x,u,u_x)]
\end{equation}
which is of the form (\ref{fuchs}). Thus an equation of the form 
(\ref{fuchsprime}) can always be reduced to one of the form (\ref{fuchs}).

The assumption made on $f$ is that it should be {\it regular} in a
sense to be defined now. It is assumed that $f$ is smooth for $t>0$ and
converges uniformly to zero on compact subsets of $(x,u,u_x)$-space as
$t\to 0$. Moreover it is assumed 
that partial derivatives of $f$ of any order with respect to $x$, $u$ and
$u_x$ satisfy the same condition. This is the definition of regularity
to be used in a smooth setting. In an analytic setting, which is where
the strongest theorem is available, it is assumed in addition that
$f$ is continuous in $t$ and analytic in the remaining arguments for
$t>0$ (cf. \cite{kichenassamy}). As to the positivity assumption
on $N$, it would be very convenient to assume that $N$ was positive
definite. Unfortunately, this often does not hold in the applications
and it is necessary to make do with a weaker condition. This can be
expressed abstractly by requiring that the matrix exponential 
$e^{tN}$ is bounded on compact sets of $x$-space for $t$ close to zero. As 
has been proved in \cite{andersson01} a sufficient condition for this is that 
the eigenvalues of $N$ are everywhere non-negative and that zero eigenvalues 
if they occur do not give rise to non-diagonal Jordan blocks.    

With these preliminaries in hand we state the main general theorem.

\vskip 10pt\noindent
{\bf Theorem}
If $f$ is regular in the analytic sense and $N$, depending analytically on
$x$, satisfies the condition that $e^{tN}$ is bounded near $t=0$ on compact 
subsets of $x$-space then the equation (\ref{fuchs}) has a unique solution 
$u$ which is regular and tends to zero as $t\to 0$.
\vskip 10pt

The definition of regularity of $u$ used here is the analogue of that for
$f$. This theorem was proved in \cite{kichenassamy}. Given that the aim
here is to construct large classes of solutions it may seem strange that
the theorem gives a unique solution. The explanation is that the equation
which the theorem is applied to arises from the original equation by a
reduction process where $u$ represents a correction to the leading order
behaviour. The leading order behaviour is given in terms of free functions
and it is these functions which give rise to the generality of the 
construction. This will be seen in more detail later. There are analogues
of the above theorem in the smooth setting but it is difficult to state
a general result which is widely applicable. The difficulty is that in
the smooth case the reduction process must give equations which are not
only in Fuchsian form but are also hyperbolic in a suitable sense.
Hyperbolicity is necessary to handle even the regular Cauchy problem.

\section{Construction of generic classes of solutions of the Einstein
equations}\label{examples}

A key test case in the application of Fuchsian methods to general
relativity is given by the Gowdy spacetimes with spatial topology
$T^3$. A heuristic analysis of the structure of singularities in
this case was given by Grubi\v{s}i\'c and Moncrief \cite{grubisic}. The
Gowdy spacetimes are solutions of the vacuum Einstein equations with
two spatial Killing vectors and an additional discrete symmetry. The
analysis of the structure of singularities in this class of spacetimes
reduces essentially to that of a pair of wave equations for two functions
$P$ and $Q$. The general solution depends on four functions of one
space variable. Its asymptotic form is
\begin{eqnarray}\label{asymptotic}
&&P(t,\theta)=k(\theta)\log t+\phi(\theta)+t^\epsilon u(t,\theta)  \\
&&Q(t,\theta)=Q_0(\theta)+t^{2k(\theta)}(\psi(\theta)+v(t,\theta))
\end{eqnarray}
for some constant $\epsilon>0$. A Fuchsian system is obtained by
writing the Einstein equations for $P$ and $Q$ in terms of $u$ and $v$.
The pair $(u,v)$ in this example plays the role of the function $u$ in
(\ref{fuchs}). The function $k$ is known as the asymptotic velocity or simply
velocity. The finding of Grubi\v{s}i\'c and Moncrief was that the above 
asymptotic form is formally consistent provided $0<k(\theta)<1$ for all
$\theta$. The inequalities imposed on $k$ constitute the low velocity
condition. In \cite{kichenassamy} it was proved that given analytic 
functions $k$, $Q_0$, $\phi$ and $\psi$ satisfying the low velocity condition 
there is a 
unique Gowdy solution for which $P$ and $Q$ have the above asymptotic form.
In the case where the low velocity condition fails it is possible to
obtain solutions depending on three free functions. The restriction which
kills the fourth free function is that $Q_0$ is required to be constant.

The low velocity condition can be related to the notion of generalized
Kasner exponents. Given a spacetime and a foliation by spacelike hypersurfaces,
let $\lambda_i$ be the eigenvalues of the second fundamental form of
the leaves of the foliation. These define scalar functions on spacetime.
If the mean curvature $\sum_i \lambda_i$ is everwhere non-zero then the
generalized Kasner exponents (GKE) are defined to be 
$p_i=\lambda_i/\sum_j \lambda_j$. They satisfy $\sum_i p_i=1$. In
spacetimes with the asymptotics described above the GKE of the foliation
by hypersurfaces of constant time converge to functions of the spatial
coordinates as the singularity is approached. These functions satisfy
the condition $\sum_i p_i^2=1$. In the Gowdy models on $T^3$ we can order
the eigenvalues, and correspondingly the GKE, in such a way that the 
indices 2 and 3 correspond to the symmetry directions. Then the low
velocity restriction is related to the condition that $p_1<0$.

There are other possible topologies for Gowdy spacetimes. They may be
defined on $S^2\times S^1$ or $S^3$. In these cases the Killing vectors
defining the symmetry have zeroes defining axes. On the axes the equations
obtained by factoring out the symmetry directions have singularities.
The possibility of generalizing the results of \cite{kichenassamy}
to Gowdy spacetimes with these other topologies was studied in \cite{stahl}.
Fuchsian techniques could be applied but in that case they did not give the
full number of free functions. It is not clear whether the restriction on
the number of free functions is an essential feature of the problem or
a failure of the technique.

If the discrete symmetry which is part of the definition of the class
of Gowdy spacetimes is dropped the class of $T^2$-symmetric
vacuum spacetimes is obtained. There is evidence to suggest that
general spacetimes of this class have oscillatory singularities 
\cite{berger01a}
and so cannot be treated by Fuchsian methods. Partially restoring the
discrete symmetry leads to the class of vacuum spacetimes with polarized
$T^2$ symmetry. In the latter class the singularity is not expected to
be oscillatory. It was shown in \cite{isenberg99} that Fuchsian techniques
can be applied to analyse the structure of the singularity in that case
in a class of spacetimes depending on the maximum number of free functions.
This class corresponds to high velocity in the sense that $p_1>0$.

The results on Gowdy spacetimes on $T^3$ have been generalized to the case
of the Einstein equations coupled to certain matter fields motivated by
string theory \cite{narita}. This is the Einstein-Maxwell-axion-dilaton 
system. The behaviour found is closely analogous to that in the vacuum case.  

All the results discussed in this section up to now concern spacetimes with
two Killing vectors and in that case the equations obtained by dividing out
the symmetry involve only one spatial variable. It should be emphasized that
a strength of Fuchsian techniques is that they do not impose a restriction
on the number of space dimensions of the system to be analysed. Thus there
is no a priori reason why they should not apply to solutions of the 
Einstein equations with less than two symmetries. The restrictions which
occur are due to the dynamics of the classes of spacetimes being 
considered near their singularities, with oscillatory behaviour obstructing
the use of Fuchsian methods.

In \cite{isenberg02} the theorem stated in the last section was applied 
to construct large classes of solutions of the vacuum Einstein equations
with $U(1)$ symmetry. In that case the equations obtained by factoring
out the symmetry are equations in two space dimensions. General solutions
with $U(1)$ symmetry are expected to show oscillatory behaviour near
their initial singularities. This has been confirmed numerically by
Berger and Moncrief \cite{berger98}. The subclass of polarized models
has singularities with monotone behaviour. These depend on two free
functions, compared to the four functions of the general case with
$U(1)$ symmetry. There is also an intermediate case, the half-polarized
case, where the solutions depend on three free functions and where
Fuchsian methods can be applied as shown in \cite{isenberg02}.

In the BKL picture singularities in general vacuum solutions in 3+1 
dimensions typically show behaviour of mixmaster type. In most cases
this is not changed by the addition of matter fields. An exception
is a linear minimally coupled scalar field. This suppresses the 
oscillations within the BKL picture. This is independent of symmetries
and indeed Fuchsian methods can be used to construct solutions of
the Einstein-scalar field system without symmetries and depending on the 
maximal number of free functions which have simple singularities whose
asymptotics can be described in great detail \cite{andersson01}.

A heuristic analysis in the spirit of BKL shows that the oscillations
near the singularity in general solutions of the vacuum Einstein
equations disappear when the spacetime dimension is at least eleven
\cite{demaret85}. A corresponding proof of the existence of large classes 
of solutions with the expected properties was given in \cite{damour}.
In that paper similar results were shown in all dimensions and in
the presence of a rather general class of field-theoretic matter models. 
The suppression of oscillatory behaviour is due to the presence
of a dilaton, a scalar field usually present in models inspired by
string theory. It is, however, important to note that a scalar field
is not in itself enough to guarantee the disappearance of oscillations.
When form fields (such as the Maxwell field) are present oscillatory
behaviour may persist.

\section{Extending the method}

The results of the last section were confined to the case where the 
functions prescribed and the solutions obtained are analytic. This
is an artificial restriction and it would be preferable to have 
results which apply to prescribed functions and solutions which
are smooth ($C^\infty$) or only of finite differentiability. Only
the smooth case will be discussed although the proofs are such that
it is clear that they would also work for functions belonging to
Sobolev spaces of sufficiently high order. The big step is that
between $C^\omega$ and $C^\infty$, since it is at this level that
it becomes important to use the hyperbolicity of the equations.

The results for Gowdy solutions on $T^2$ described in the last section
were generalized to the case of $C^\infty$ solutions in \cite{rendall00}.
The methods of proof used are likely to be of much wider applicability
but significant additional work is still required in any particular
example. The only additional case in which this work has been done so 
far is that of Gowdy spacetimes with spherical topology \cite{stahl}.

There is older work in which equations of Fuchsian type were used to
determine the structure of singularities in solutions of the Einstein
equations without symmetry. These solutions are subject to another
kind of restriction; they have isotropic singularities. The notion
of an isotropic singularity arose in the context of Penrose's Weyl
curvature hypothesis. Penrose proposed that initial singularities
of solutions of the Einstein equations should have a special
structure corresponding in some sense to a state of low entropy.
This provides a selection criterion for initial data for cosmological
models. With this motivation in mind it is interesting to know how
many solutions of the Einstein equations there are with singularities
satisfying this condition. The first general results on this question
were developed in \cite{newman93}. The required existence proofs were
provided in \cite{claudel}. They make heavy use of semigroup theory.

The results of \cite{newman93} concerned solutions of the Einstein
equations coupled to a perfect fluid with equation of state 
$p=\frac13\rho$. They were generalized to a fluid with
the more general equation of state $p=k\rho$ in \cite{anguige99a}.
In both cases the theorems obtained show that the solutions having
an isotropic singularity depend on half the number of free 
functions allowed by general solutions of the Einstein equations.
In \cite{anguige00a} theorems were proved where the Euler equations
are replaced by the Vlasov equation for collisionless matter.
Note that the theorem of Section \ref{kr} does not apply directly to the 
Einstein-Vlasov system since this is a system of integrodifferential
equations instead of a system of differential equations as in the
hypotheses of that theorem. The picture of isotropic singularities 
obtained in the case of collisionless matter is quite different from that 
for a fluid. There is much more freedom to give initial data. The initial
phase space density of particles can be prescribed almost without
restriction. This discrepancy suggested looking at the case of matter 
described by the Boltzmann equation, which is in a sense intermediate 
between the Euler and Vlasov descriptions. It turns out that the 
behaviour of the collision cross section for large momenta plays a 
crucial role. If the cross section grows fast enough in the limit
of large momenta the initial phase space density is forced to
be in equilibrium which effectively reduces this case to the fluid
case. If, on the other hand, the collision cross section has slow 
growth the freedom in the initial phase space density is as great
as in the Vlasov case. This has been shown on the level of formal
expansions by Tod \cite{tod} although no existence proofs are yet
available. 

At several points in the above we have used function counting as
a criterion for the generality of solutions. This is not very
satisfactory and has been used up to now since it is the only
thing which can presently be proved. What is desirable is to
have theorems which say that the solutions having a particular
asymptotic form near their singularities include all solutions 
arising from a non-empty open set of initial data on a regular
Cauchy surface. It would be even better if this open set could
also be described so as to see what kind of restriction on 
initial data it really represents. So far these goals have only
been reached in one case, namely in the work of Ringstr\"om
\cite{ringstrom02} concerning Gowdy solutions on $T^3$. 

One obstruction to the application of Fuchsian methods has already
been described. This is the fact that the solutions may show a
complicated oscillatory behaviour near the singularity. It is
not the only obstruction. Another phenomenon which arises naturally
within the BKL picture is that small-scale spatial structures can be 
formed as the singularity is approached. In situations where
Fuchsian techniques can be applied it is typical that spatial
derivatives of important geometrical quantities blow up at a
rate which is comparable to the rate with which the undifferentiated
quantities blow up as the singularity is approached. In the 
phenomenon of formation of small-scale spatial structure this
property of the spatial derivatives no longer holds. The simplest
instance is the formation of spikes in Gowdy solutions \cite{rendall01}.
In that paper families of examples were constructed which depend on
the maximum number of free functions. Thus in the sense of function 
counting Gowdy solutions with spikes as common as those without.
The development of spikes is associated with the presence of velocities
greater than one. The low velocity condition in the Fuchsian existence
theorems for Gowdy solutions prevents the occurrence of spikes. It
is to be expected on the basis of the BKL picture that the formation of 
small-scale structure will occur more generally and there is 
numerical evidence supporting this \cite{berger98}.

It has already been mentioned that the possible applications of
Fuchsian methods to field-theoretic matter models have been worked
out in some generality in \cite{damour}. Symmetry assumptions
can lead to further tractable cases and some of these were looked
at in \cite{narita}. Much less has been done for phenomenological
matter models such as perfect fluids and kinetic theory. Apart
from the work on isotropic singularities the only available results
are those of Anguige \cite{anguige00b}, \cite{anguige00c} on certain
solutions of the Einstein-Euler system with two spacelike Killing
vectors. What happens for spacetimes without symmetry with a scalar 
field and a fluid or kinetic matter?

There is an application of Fuchsian methods to the asymptotics of
expanding cosmological models which can be envisaged. Cosmological
models with positive cosmological constant $\Lambda$ often exhibit an
inflationary phase of exponential expansion. At least formally 
the asymptotics obtained seems to be such as to be amenable to
a Fuchsian treatment. As yet there is no analysis of this kind
in the literature. The idea of a mathematical similarity between
certain spacetime singularities and inflationary expanding phases
of cosmological models opens up the perspective of an interesting
exchange of information between apparently unrelated areas of
research. A formal analysis of the asymptotics of expanding cosmological
models with $\Lambda>0$ has been given in \cite{starobinsky}. A
corresponding treatment of the case of power-law inflation, where
the cosmological constant is replaced by a scalar field with
exponential potential, can be found in \cite{muller}.

To finish, we mention a different kind of application of Fuchsian
methods in relativity. In this case no spacetime singularities are
involved. Instead the singularity in the equations arises from a
coordinate singularity due to the use of polar coordinates. Consider
a spherically symmetric solution of the Einstein equations coupled
to some matter fields. If the solution is in addition static the
Einstein equations reduce to a system of ordinary differential equations
in a radial variable $r$. It has a singularity at the centre of symmetry,
$r=0$. In the case that the matter is a fluid existence theorems were
obtained in \cite{rendall91}. The proofs were based on an existence 
theorem for singular ODE which, with hindsight, belongs to the class
of Fuchsian equations. Corresponding results in the cases of
collisionless matter and elastic solids were proved in \cite{rein}
and \cite{park} respectively. In the former case, integrodifferential
equations occur, the theorem of \cite{rendall91} is not applicable, and
existence has to be proved by hand. The same existence theorem for 
singular ODE has been applied in \cite{noundjeu} to analyse the 
constraint equations doe the Einstein-Vlasov-Maxwell system in the 
spherically symmetric case,

\end{document}